\documentclass{amos}

\usepackage{hyperref}
\usepackage{caption}
\usepackage{subcaption}
\usepackage{float}
\usepackage{graphicx}
\usepackage{amsbsy}
\usepackage{amsfonts}
\usepackage{amsmath}
\usepackage{multirow}
\usepackage{booktabs}
\usepackage[table]{xcolor}

\title{\TitleFont Transformer-based Atmospheric Density Forecasting}

\author{ \footnote{Ph.D. Student, Department of Aeronautics and Astronautics. E-mail: jbriden@mit.edu}}

\author{Julia Briden \footnote{Ph.D. Student, Department of Aeronautics and Astronautics. E-mail: jbriden@mit.edu}, \; Peng Mun Siew \footnote{Postdoctoral Associate, Department of Aeronautics and Astronautics. E-mail: siewpm@mit.edu}, 
\\ Massachusetts Institute of Technology \and \vspace{2 pt}
Victor Rodriguez-Fernandez \footnote{Associate Professor, Department of Computer Systems Engineering. E-mail: victor.rfernandez@upm.es}, \\ Universidad Politécnica de Madrid \and 
\vspace{-7 pt}
Richard Linares\footnote{Associate Professor, Department of Aeronautics and Astronautics. E-mail: linaresr@mit.edu} \\ Massachusetts Institute of Technology}

\date{}

\begin{document} 

\maketitle

\begin{abstract}
	\normalsize
As the peak of the solar cycle approaches in 2025 and the ability of a single geomagnetic storm to significantly alter the orbit of Resident Space Objects (RSOs), techniques for atmospheric density forecasting are vital for space situational awareness. While linear data-driven methods, such as dynamic mode decomposition with control (DMDc), have been used previously for forecasting atmospheric density, deep learning-based forecasting has the ability to capture nonlinearities in data. By learning multiple layer weights from historical atmospheric density data, long-term dependencies in the dataset are captured in the mapping between the current atmospheric density state and control input to the atmospheric density state at the next timestep. This work improves upon previous linear propagation methods for atmospheric density forecasting, by developing a nonlinear transformer-based architecture for atmospheric density forecasting. Empirical NRLMSISE-00 and JB2008, as well as physics-based TIEGCM atmospheric density models are compared for forecasting with DMDc and with the transformer-based propagator.
	
\end{abstract}

\section{Introduction}
	\label{sec:intro}

Thermospheric mass density serves as the largest source of uncertainty for low Earth orbit (LEO) satellite orbit prediction. This uncertainty is largely due to fluctuations in solar and geomagnetic activity that can occur on the order of hours. Solar and geomagnetic storms take place when charged particles, usually from coronal mass ejections, travel to Earth’s atmosphere and increase atmospheric heating and transient solar wind activity. These storms are often marked by high nonlinearities in the evolution of atmospheric density over time and, therefore, prove to be difficult to predict with even the most advanced forecasting techniques. With the ability of a single geomagnetic storm to significantly alter the orbit of satellites, techniques for atmospheric density forecasting are vital for space situational awareness.

Current forecasting methods use physics-based atmospheric density models, such as the Global Ionosphere-Thermosphere Model (GITM) and the Thermosphere-Ionosphere-Electrodynamics General Circulation Model (TIE-GCM), which solve the full continuity, energy, and momentum equations required for the propagation of atmospheric density. While these models are ideal for short-term forecasting, the computational cost required to run physics-based models exceeds computing capabilities for most real-time applications. Conversely, empirical models, including NRLMSISE-00 and JB2008, are fast to evaluate, since they are derived only from measurements, such as total mass density, temperature, and oxygen number density. Unfortunately, empirical models cannot translate their computational efficiency to atmospheric density prediction, since they lack the underlying dynamics required for forecasting. Recent work in reduced-order modeling (ROM) and dynamic mode decomposition with control (DMDc) has addressed this gap in computationally efficient atmospheric density prediction methods; ROMs use proper orthogonal decomposition (POD) or convolutional autoencoder-based machine learning (ML) to reduce the dimensionality of physics-based models and generate reduced-order snapshots of empirical models to enable atmospheric density propagation. To propagate a reduced-order atmospheric density state forward in time, DMDc models atmospheric density as a linear dynamical system with a control input, where the dynamics and input matrices are estimated from a dataset of state snapshots. While this approach captures the nominal atmospheric density dynamics relatively well, DMDc often fails when mildly nonlinear conditions occur \cite{wu2021}.

The effectiveness of using a machine learning approach for modeling the nonlinear dynamics in atmospheric density has been assessed by Turner et al. \cite{Turner2020}. By utilizing a deep feedforward neural network (NN) for atmospheric density forecasting, an error reduction of over 99 \%, when compared to DMDc, was achieved. While effective for short-term forecasts, prediction performance was hindered by the inability of the feedforward NN to optimize with data from previous time steps. Moreover, the challenge of atmospheric density forecasting during a significant space weather event requires an algorithm that can capture long-term dependencies in the dataset to prevent compounding propagation errors. The NN must handle large inputs for the atmospheric density state and only focus on the relevant part of the input for accurate forecasting. With an attention component, when compared to Recurrent Neural Networks (RNNs), transformers train on sequential data in less time and with longer inputs by using a mechanism known as attention \cite{wen2023}. By passing all hidden states, derived from the encoded sequential atmospheric density reduced-order states and space weather input, to the transformer NN, the transformer can focus its attention on only the most relevant hidden states. Since the occurrence of a solar or geomagnetic storm represents an anomaly in the atmospheric density dynamical system, the transformer’s ability to focus attention on only the relevant propagation dynamics is imperative for fast and accurate atmospheric density forecasting.

After reducing the dimensionality of the current atmospheric density state using a POD or ML ROM, the transformer-based atmospheric density forecasting algorithm takes the current and past reduced-order states and the current space weather indices as an input to the transformer’s encoder network and generates a new reduced-order atmospheric density state at the next time step. Where the encoder generates a set of embeddings for the control input and the time series reduced-order density states, then outputs the predicted reduced-order density states using the encoder’s embeddings. The training process includes dataset splits for low, medium, and high levels of space weather activity to improve generalization performance. The final transformer propagation model provides a mapping between time-series reduced-order states and control input to the next predicted state, serving as a surrogate dynamical system.

Empirical NRLMSISE-00 POD ROM, JB2008 POD ROM, and JB 2008 ML ROM, as well as physics-based TIEGCM POD ROM atmospheric density models are compared for forecasting with DMDc and with the transformer-based propagator. With almost 5,500 active satellites currently in orbit and the predicted launch of an additional 58,000 by 2030, forecasting the nonlinear dynamics of space weather-induced changes in atmospheric density are essential for Resident Space Object (RSO) deorbit prediction and collision avoidance.

\section{Theory}
\label{sec:Theory}

To achieve a data-driven model of the global atmospheric density dynamical system, two steps are completed; first, the full-order atmospheric density is reduced in space, either using proper orthogonal decomposition (POD) or a machine learning-based (ML) convolutional autoencoder. Then the reduced-order snapshot is propagated forward using dynamic mode decomposition with control (DMDc) or a transformer neural network. In the following section, POD and ML spatial reduction theory, as well as DMDc forecasting, will be reviewed. For additional information on these modeling methods, refer to \cite{briden2022, clark2023reduced, Mehta2017, Loffe2015, DeCarlo1989}. Then transformer-based forecasting for time series data is covered.

\subsection{Reduced-Order Modeling}
\label{subsec:TheoryROMs}
In the following subsection, we cover the development of the various reduced-order models used in the study.
\subsubsection{Proper Orthogonal Decomposition}
\label{subsub:TheoryROMPOD}

    The variation in atmospheric mass density is defined as $\tilde{\mathbf{x}}(\mathbf{s},t) = \mathbf{x}(\mathbf{s},t) - \bar{\mathbf{x}}(\mathbf{s})$. By subtracting the mean atmospheric density, $\bar{\mathbf{x}}(\mathbf{s})$ for a spatial grid and time index, from the true atmospheric mass density, \textbf{x}. Then, the reduced-order state, \textbf{z}, can be constructed using the first r POD modes from the singular value decomposition, $\mathbf{U_{r}}$, resulting in Eqn. \ref{eq:reducedState}:
    
    \begin{equation}
    \mathbf{z} =  \mathbf{U}_{r}^{-1} \tilde{\mathbf{x}} = \mathbf{U}_{r}^{T} \tilde{\mathbf{x}}.
    \label{eq:reducedState}
    \end{equation}
    
    When the full-dimensional state is desired, the first r POD modes can be multiplied by the reduced-order state while adding back the mean atmospheric density:
    
    \begin{equation}
    \mathbf{x}(\mathbf{s},t) \approx \mathbf{U}_{r}(\mathbf{s}) \mathbf{z}(t) + \bar{\mathbf{x}}(\mathbf{s}).
    \label{eq:projection}
    \end{equation}
    
    For additional information on the applications of the POD algorithm for thermospheric mass density prediction see Mehta and Linares (2017) \cite{Mehta2017}.

\subsubsection{Machine Learning-based Dimensional Reduction}
\label{subsub:TheoryROMML}

    The other dimensionality reduction technique employed in this work is the undercomplete convolutional autoencoder neural network \cite{Goodfellow2016}. Where the term undercomplete defines the encoded dimension as less than the input dimension. The encoder, $\mathbb{F}$, maps the input vector to a reduced-order vector:

    \begin{equation}
    \mathbb{F} : X \rightarrow V.
    \label{eq:autoencoderEncode}
    \end{equation}
    
    While the decoder, maps a reduced-order vector to its full-dimensional form:

    \begin{equation}
    \mathbb{G} : V \rightarrow X'.
    \label{eq:autoencoderDecode}
    \end{equation}
    
    To assess the accuracy of these learned mappings, a mean squared error loss function is used for this work:

    \begin{equation}
    \mathcal{L}(X,X') = \frac{||X - X'||^{2}}{d}.
    \label{eq:lossFunction}
    \end{equation}

    Where $d$ defines the dimension of the input vector and its reconstructed dimension. The convolutional autoencoder utilizes the 2D convolution for the output, $g$, and the input, $x$:
    
    \begin{equation}
     g_{i,j,k,c} = \sigma (w_{c}^{T} x_{i,j,k,c}).
    \label{eq:2dConvolution}
    \end{equation}
    
    Where $g_{i,j,k,c}$ is a data in the channel output and c is the filter. While training, proximal policy optimization (PPO) was utilized for parameter (weights and biases) tuning. See Loffe et al. (2015) for additional information on the convolutional autoencoder neural network and Briden et al. (2022) for the full neural network architecture used in this work \cite{Loffe2015, briden2022}.

    Compared to the linear POD, the ML dimensionality reduction approach has the benefit of capturing nonlinear features in the dataset. This benefit comes at a risk for overfitting to training data, which is mitigated by including dropout in the convolutional autoencoder model used in this work.

\subsubsection{Dynamic Mode Decomposition with Control}
\label{subsub:TheoryROMDMDc}

    Once either POD or ML are used to reduce the dimensionality of the atmospheric density state, to achieve forecasting for a later time, a propagation model is formulated. In previous work, Dynamic Mode Decomposition with control (DMDc) was utilized to learn a linear dynamical system model from data \cite{briden2022}. The future state at time index $k+1$ is constructed using Eqn. \ref{eq:linearDynamics}:
    
    \begin{equation}
     \mathbf{z}_{k+1} = \mathbf{A} \mathbf{z}_{k} + \mathbf{B} \mathbf{u}_{k}
    \label{eq:linearDynamics}
    \end{equation}

    where $\mathbf{z}_{k}$ is the reduced-order state at time index k, as defined in Eqn. \ref{eq:reducedState} or Eqn. \ref{eq:autoencoderEncode}), and $\mathbf{u}$ is the set of space weather indices. The matrices, \textbf{A} and \textbf{B} are estimated using the method of least squares given the time-shifted snapshot matrices for a fixed timestep, $T$. To convert the discrete-time dynamics into continuous-time dynamics, the relation from DeCarlo (1989) is utilized \cite{DeCarlo1989}.

    \begin{equation}
    \begin{bmatrix}
    \mathbf{A}_{c} & \mathbf{B}_{c}\\
    \mathbf{0} & \mathbf{0}
    \end{bmatrix} 
    = log(
    \begin{bmatrix}
    \mathbf{A} & \mathbf{B}\\ \mathbf{0} & \mathbf{I}
    \end{bmatrix}) 
    / T.
    \label{eq:matrices}
    \end{equation}

    After propagating for the desired time length, either Eqn. \ref{eq:projection}, for POD, or Eqn. \ref{eq:autoencoderDecode}, for the Autoencoder NN, can be used to project the predicted reduced-order atmospheric density to the full dimensional state.

    Similar to the trade-offs between POD and ML for dimensionality reduction, there exists use cases for development of a propagation method which captures the nonlinearities in the atmospheric density dynamics. Furthermore, noise sensitivity serves as a prominent challenge in DMD \cite{wu2021}. This challenge can inhibit DMDc propagation for long time periods. By constructing a transformer-based NN propagator, current challenges in robust nonlinear propagation during significant space weather events can be addressed.

\subsubsection{Transformer Propagator}
\label{subsub:TheoryTransformer}

    To construct the transformer forecaster, the channel-independent patch time series Transformer (PatchTST) architecture was utilized \cite{nie2023}. By using a patching and channel-independence design, local semantic information in the time series is preserved and model complexity can be reduced.

    Due to channel-independence, the transformer forecaster diverges from the DMDc linear dynamical system model; both reduced-order atmospheric density states and the space weather inputs are combined into one state snapshot vector. While this architecture enables prediction of both atmospheric density and space weather indices, the architecture no longer assumes atmospheric density's dependence on the space weather input. Future work will focus on expanding the Transformer propagator to include a graph neural network to learn the cross-channel relationships between space weather indices and atmospheric density. Although only the channel-independent transformer propagation technique is assessed in this work, this simplified framework serves to improve robustness to noise; channel independence mitigates the possibility of projecting noise from one input channel into another input channel. A significant improvement from the noise-sensitive DMDc.

    To propagate the state snapshot forward in time, the transformer backbone, $\mathbb{H}$, maps the reduced-order state vector, $\mathbf{z}_{1:L} = (z_1, ..., z_L)$, to the predicted reduced-order state, $\hat{\mathbf{z}}_{L+1:L+T} = (\hat{z}_{L+1}, ..., \hat{z}_{L+T})$:

    \begin{equation}
    \mathbb{H} : Z \rightarrow \hat{Z}.
    \label{eq:transformerMap}
    \end{equation}

    Where $L$ is look-back window length to provide predictions based on and $T$ is the forecast window length for the transformer NN to output. Rather than predicting the next density snapshot index from the previous density snapshot index at a fixed rate, as DMDc does, the PatchTST architecture allows for an increase in information utilized to generate a prediction and a longer prediction window for one forward pass, resulting in potentially significant accuracy improvements, as well as fast predictions for long propagation windows.

    Through patching, each input time series is divided into patches that can overlap or not overlap. The patch length is denoted as $P$ and the stride, or non-overlapping region between two patches, is $S$. Then the number of patches is $N = \lfloor \frac{L-P}{S} \rfloor + 2$. Patching reduces memory usage and computational complexity of the attention map quadratically by a factor of $S$ \cite{nie2023}. Since this allows the model to view longer historical data, forecasting performance can be significantly improved.

    The encoder used in this work is a vanilla Transformer encoder that maps the observed signals to the latent representations \cite{nie2023}. The patches are mapped to the Transformer latent space of dimension $D$ via a trainable linear projection $W_p \in \mathbb{R}^{D \times P}$, and a learnable additive position encoding $W_{pos} \in \mathbb{R}^{D \times N}$ is applied to monitor the temporal order of patches:

    \begin{equation}
    z_d^{(i)} = W_p z_p^{(i)} + W_{pos}
    \end{equation}
    
    where $z_d^{(i)} \in \mathbb{R}^{D \times N}$ denotes the input that will be fed into the Transformer encoder. Then each head $h = 1, ..., H$ in multi-head attention will transform them into query matrices $Q_h^{(i)} = (z_d^{(i)})^T W_{Q_h}$, key matrices $K_h^{(i)} = (z_d^{(i)})^T W_{K_h}$, and value matrices $V_h^{(i)} = (z_d^{(i)})^T W_{V_h}$, where $W_{Q_h}, W_{K_h} \in \mathbb{R}^{D \times d_k}$ and $W_{V_h} \in \mathbb{R}^{D \times D}$. After that, a scaled production is used for getting attention output $O_h^{(i)} \in \mathbb{R}^{D \times N}$:
    
    \begin{equation}
    (O_h^{(i)})^T = \text{Attention}(Q_h^{(i)}, K_h^{(i)}, V_h^{(i)}) = \text{Softmax}\left(\frac{Q_h^{(i)} (K_h^{(i)})^T}{\sqrt{d_k}}\right) V_h^{(i)}
    \end{equation}
    
    The multi-head attention block also includes BatchNorm1 layers and a feedforward network with residual connections \cite{Loffe2015}. Afterwards, it generates the representation denoted as $z^{(i)} \in \mathbb{R}^{D \times N}$. Finally, a flatten layer with a linear head is used to obtain the prediction result $\hat{z}^{(i)} = (\hat{z}_{L+1}^{(i)}, ..., \hat{z}_{L+T}^{(i)}) \in \mathbb{R}^{1 \times T}$.

    As in Section \ref{subsub:TheoryROMML}, MSE loss (Eqn. \ref{eq:lossFunction}) is used to measure prediction accuracy. Finally, instance normalization is utilized in the transformer architecture to mitigate distribution shift between training and test data. For additional details on the PatchTST architecture, see Nie et al. (2023) \cite{nie2023, tsai}.

\section{Methods}
\label{sec:Methods}

To assess the forecasting ability of the transformer NN propagator, test datasets consisting of high, medium, and low space weather activity were constructed. POD ROMs for the JB2008, NRLMSISE-00, and TIEGCM atmospheric density models and an ML ROM for the JB2008 atmospheric density model from Briden et al. (2022), along with their associated space weather inputs, are the models used in this study \cite{briden2022}. The new transformer propagation method is compared to DMDc.

\subsection{Space Weather Inputs}

Space weather activity in the form of space weather drivers were used for the division of training, validation, and test data for the transformer propagator, as well as inputs for the DMDc propagator. Space weather indices serve as a proxy for levels of space weather activity. Table \ref{linearInput} shows the list of space weather indices used by the atmospheric density models in this work.

\begin{table}[htb!]
    \caption{Space Weather Inputs}
	\centering
	\begin{tabular}{| l | c | c |}
		\hline 
		\textbf{Index}       & \textbf{Description} & \textbf{Models Used}  \\
		\hline
		\multirow{ 2}{*}{$F_{10.7}$} & Solar radio noise flux at a  & JB2008, POD TIE-GCM ROM,   \\
		& wavelength of 10.7 cm & NRLMSISE-00\\
		\hline
		\multirow{ 2}{*}{$S_{10}$} & Activity indicator of the integrated  & \multirow{ 2}{*}{JB2008} \\
		& 26–34 nm solar irradiance  & \\
		\hline
		\multirow{1}{*}{$M_{10}$} & Modified daily Mg II core-to-wing ratio  & JB2008  \\
		\hline
		\multirow{ 2}{*}{DSTDTC} & Temperature change calculated from & \multirow{ 2}{*}{JB2008}    \\
		& the Disturbance Storm Time (DST) Index  & \\ 
		\hline
		\multirow{ 2}{*}{kp} & 3-hour-range standardized  &  \multirow{ 2}{*}{POD TIE-GCM ROM}	\\
		& quasi-logarithmic magnetic activity  & \\ 
		\hline
		\multirow{ 2}{*}{ap} & 3-hourly magnetic activity index  & \multirow{ 2}{*}{NRLMSISE-00} \\
		& derived from kp  & \\ 
		\hline
	\end{tabular}
    \label{linearInput}
    \end{table}
    
    For ROM propagation with DMDc, additional future (next-hour) and nonlinear space weather indices were used and are shown in Table \ref{tab:sw_nl_input}. The JB2008 model uses time-lag solar flux indices, each with a different time lag. For the JB2008 model, future indices are defined as next-day values for each of the time-lag solar flux indices. Where F10 and S10 have a one day lag, M10 has a two day lag, and Y10 has a 5 day lag. These delays exist in the F10 index for the NRLMSISE model as well.  As demonstrated in \cite{Gondelach2020}, the usage of future and nonlinear space weather indices improve the accuracy of the DMDc prediction. These nonlinear space weather indices were used as inputs for all ROM models.
    
    \begin{table}[hbt!]
    \caption{Additional Space Weather Inputs Used for the DMDc Models}
    \label{tab:sw_nl_input}
    \centering
    \begin{tabular}{|l |c |c |c|}
    \hline
     \textbf{ROM Model}  & \textbf{Standard input} & \textbf{Future input} & \textbf{Nonlinear input}  \\
    \hline
      \multirow{ 2}{*}{\textbf{JB2008}}  & doy, hr, $F_{10.7}$, $\overline{F}_{10.7}$, $S_{10}$, $\overline{S}_{10}$, $M_{10}$, $\overline{M}_{10}$, & $F_{10.7}$, $S_{10}$, $M_{10}$ & $DSTDTC^2$, \\
      & $Y_{10}$, $\overline{Y}_{10}$,$DSTDTC$, $GMST$, $\alpha_{sun}$, $\delta_{sun}$ & $Y_{10}$, $DSTDTC$ & $F_{10.7} \cdot DSTDTC$ \\
      \hline
      \textbf{TIE-GCM}  & doy, hr, $F_{10.7}$, $\overline{F}_{10.7}$, $kp$   & $F_{10.7}$, $kp$ & $kp^2$, $kp\cdot F_{10.7}$\\
    \hline
    \multicolumn{4}{l}{Note. doy = day of year; hr = hour in UTC; GMST = Greenwich Mean Sidereal Time. Overbars indicate the} \\
    \multicolumn{4}{l}{81-day average. Nonlinear inputs are constructed using both current inputs and future inputs.}
    \end{tabular}
    \end{table}

\subsection{Transformer Propagator Training}
\label{subsec:MethodsTraining}

\begin{figure}[htb!]
    \includegraphics[width=0.5\textwidth]{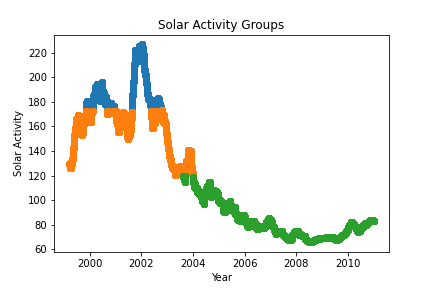}
    \centering
    \caption{Grouped solar activity levels for the F10B solar index. Where blue is high solar activity, orange is medium solar activity, and green is low solar activity.}
    \label{fig:solar_activity}
    \end{figure}

    To develop training, validation, and test data, POD ROMs for the JB2008, NRLMSISE, and TIEGCM atmospheric density models and an ML ROM for the JB2008 atmospheric density model were split based on the level of solar activity, as shown in Figure \ref{fig:solar_activity}. 
    
    The F10B solar index was used for JB2008 and the F10a solar index was used for NRLMSISE and TIEGCM solar activity grouping. Both of these indices correspond to 81-day centered averages for the solar radio noise flux at a wavelength of 10.7 cm.
    
\begin{figure}[htb!]
    \includegraphics[width=0.5\textwidth]{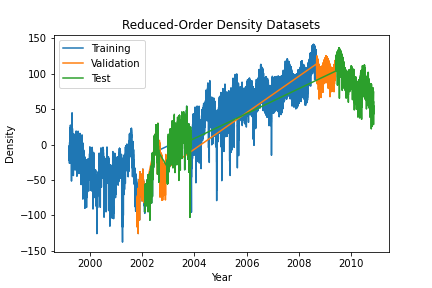}
    \centering
    \caption{Combined training, validation, and test data for the first atmospheric density POD reduced-order state.}
    \label{fig:dataset}
    \end{figure}

    Then each of these three space weather activity datasets were split for training, validation, and test data. The splits included 20\% test data and 10\% validation data. A one week forecast horizon and a 42 day look-back window were chosen for the transformer propagator. After each of the three datasets (high, medium, and low solar activity) were divided into training, validation, and test data, a combined dataset was created by concatenating the forecasting splits for all levels of solar activity. Figure \ref{fig:dataset} shows the first JB2008 POD reduced-order state for the corresponding training, validation, and test splits.

    While training, validation, and test datasets are not sequential in time, this structure allows for the training process to have representative samples from a range of solar activity levels, while having non-overlapping test samples to evaluate performance for a variety of test samples. Since only one solar cycle of model data is available, this split structure is required for maximum forecasting performance. Furthermore, none of the model variables indicate the timestamp of the sample. Even though this split structure does not represent a sequential use of the model, time information is not leaked when mixing future and past samples between training, validation and test.

    The model architecture for the JB2008 POD transformer propagator is included in Table \ref{table:NN structure}. The transformer propagator for each model, including POD ROMs for the JB2008, NRLMSISE, and TIEGCM atmospheric density models and an ML ROM for the JB2008 atmospheric density model, have the same architecture: 1 encoder layer, 2 heads, 20-dimensional model, 28-dimensional fully connected netork, no attention dropout, 0.2 dropout applied to all linear layers in the encoder except q, k, and v projections, a patch length of 9, and a stride length of 1.
    
    \begin{table}[ht]
    \centering
    \begin{tabular}{|l|l|l|l|}
    \hline
    \textbf{Layer (type)} & \textbf{Output Shape} & \textbf{Param \#} & \textbf{Trainable} \\
    \hline
    RevIN & 1 x 34 x 168 & 68 & True \\
    \hline
    ReplicationPad1d & 1 x 34 x 1009 & - & - \\
    \hline
    Unfold & 1 x 9 x 1001 & - & - \\
    \hline
    Linear & 1 x 34 x 1001 x 20 & 200 & True \\
    \hline
    Dropout & - & - & - \\
    \hline
    Linear & 1 x 34 x 1001 x 20 & 420 & True \\
    \hline
    Linear & 1 x 34 x 1001 x 20 & 420 & True \\
    \hline
    Linear & 1 x 34 x 1001 x 20 & 420 & True \\
    \hline
    Dropout & - & - & - \\
    \hline
    Linear & 1 x 34 x 1001 x 20 & 420 & True \\
    \hline
    Dropout & - & - & - \\
    \hline
    Dropout & - & - & - \\
    \hline
    Transpose & 1 x 20 x 1001 & - & - \\
    \hline
    BatchNorm1d & 1 x 20 x 1001 & 40 & True \\
    \hline
    Transpose & 1 x 1001 x 20 & - & - \\
    \hline
    Linear & 1 x 1001 x 28 & 588 & True \\
    \hline
    GELU & - & - & - \\
    \hline
    Dropout & - & - & - \\
    \hline
    Linear & 1 x 1001 x 20 & 580 & True \\
    \hline
    Dropout & - & - & - \\
    \hline
    Transpose & 1 x 20 x 1001 & - & - \\
    \hline
    BatchNorm1d & 1 x 20 x 1001 & 40 & True \\
    \hline
    Transpose & 1 x 1001 x 20 & - & - \\
    \hline
    Flatten & 1 x 34 x 20020 & - & - \\
    \hline
    Linear & 1 x 34 x 168 & 3363528 & True \\
    \hline
    \end{tabular}
    \caption{PatchTST Model Structure: 3,366,724 total parameters (Input shape: 1 x 34 x 1008).}
    \label{table:NN structure}
    \end{table}

Both the DMDc ROMs and transformer ROMs were trained on the same dataset, shown in Figure \ref{fig:dataset} in blue, which includes data from low, medium, and high levels of space weather activity. Then for testing, the DMDc and transformer propagators were both evaluated on different levels of space weather activity to determine model performance under representative space weather conditions.

\section{Results}
\label{sec:Results}

To evaluate the performance of the transformer forecasting approach, compared to the DMDc forecasting approach, the mean-squared-error (MSE) and mean-absolute-error (MAE) of model predictions were computed for 1 week prediction windows on the low, medium, and high solar activity test datasets. Plots were generated to show the propagation of MSE over 1 week periods and the DSTDTC and Kp weather inputs are plotted to show changes in space weather activity. Instead of plotting F10, DSTDTC was chosen, since it proved to be a better indicator of short-term space weather activity. The DMDc JB2008 models use DSTDTC as an input and the DMDc TIEGCM model uses Kp as an input. NRLMSISE uses ap as a input, which is derived from Kp.

\begin{figure}[htb!]
\includegraphics[width=1\textwidth]{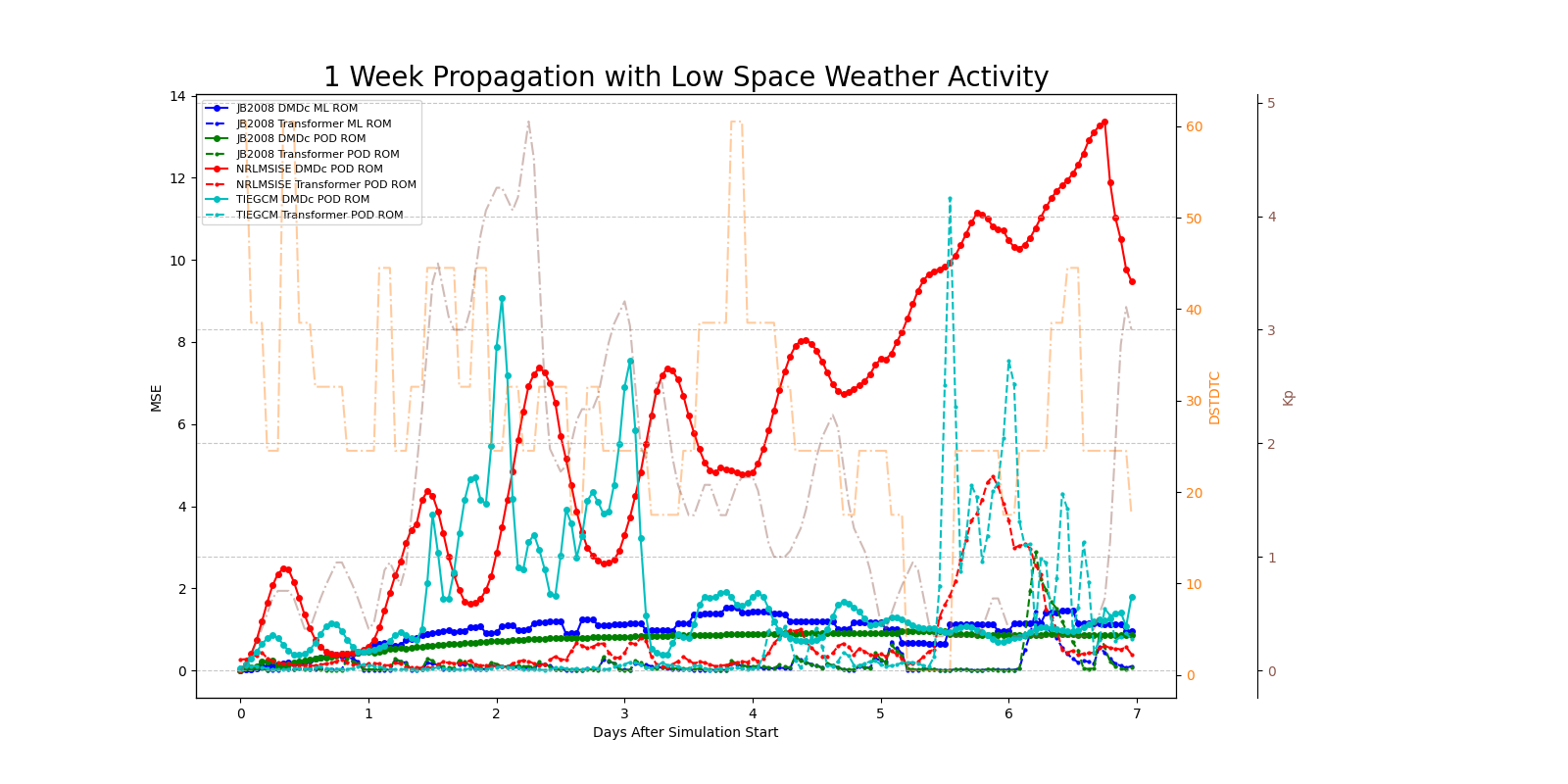}
\centering
\caption{MSE for one week propagation on low space weather activity test data.}
\label{fig:propagationlow}
\end{figure}

\begin{figure}[htb!]
\includegraphics[width=1\textwidth]{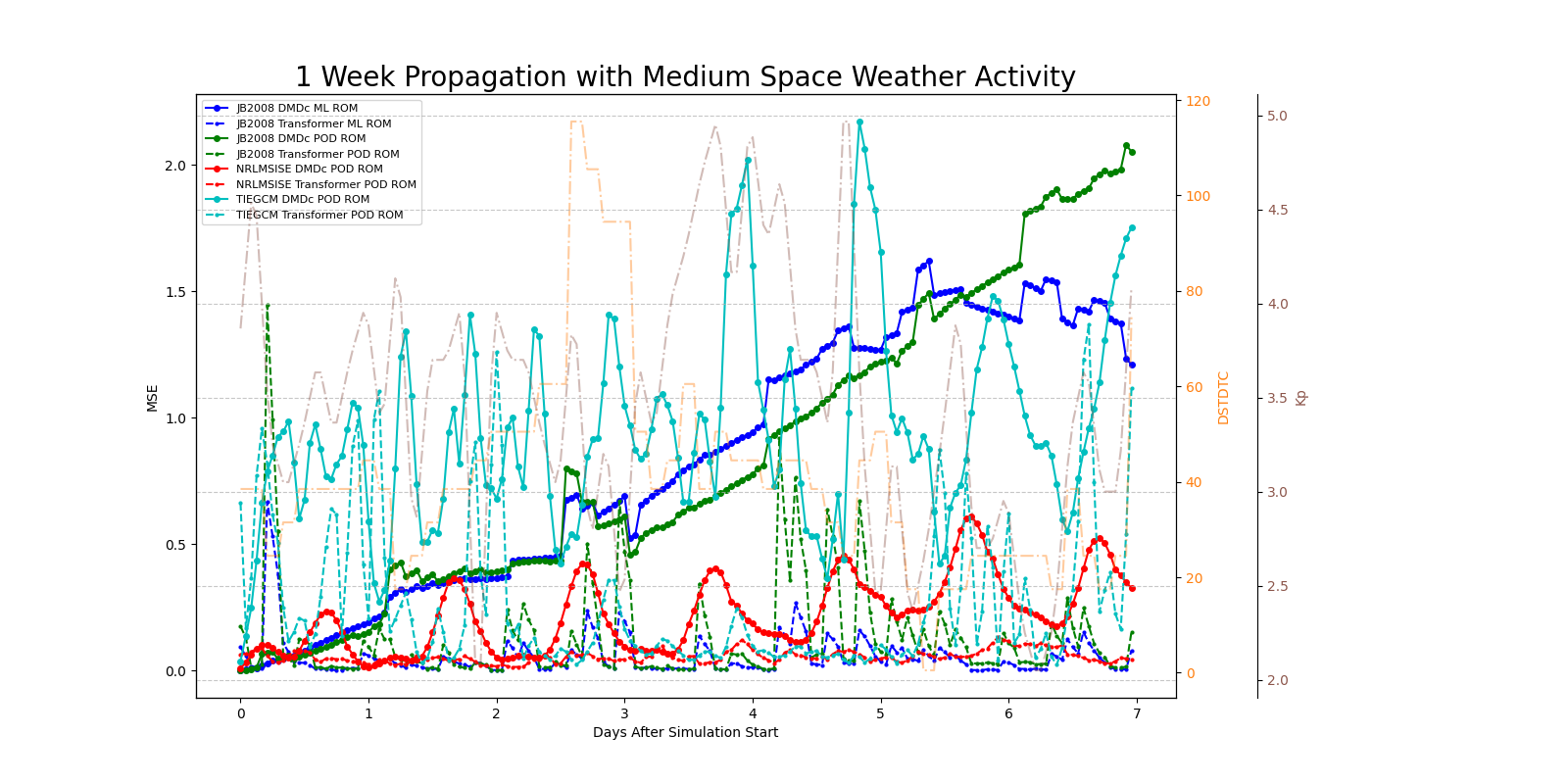}
\centering
\caption{MSE for one week propagation on medium space weather activity test data.}
\label{fig:propagationmedium}
\end{figure}

\begin{figure}[htb!]
\includegraphics[width=1\textwidth]{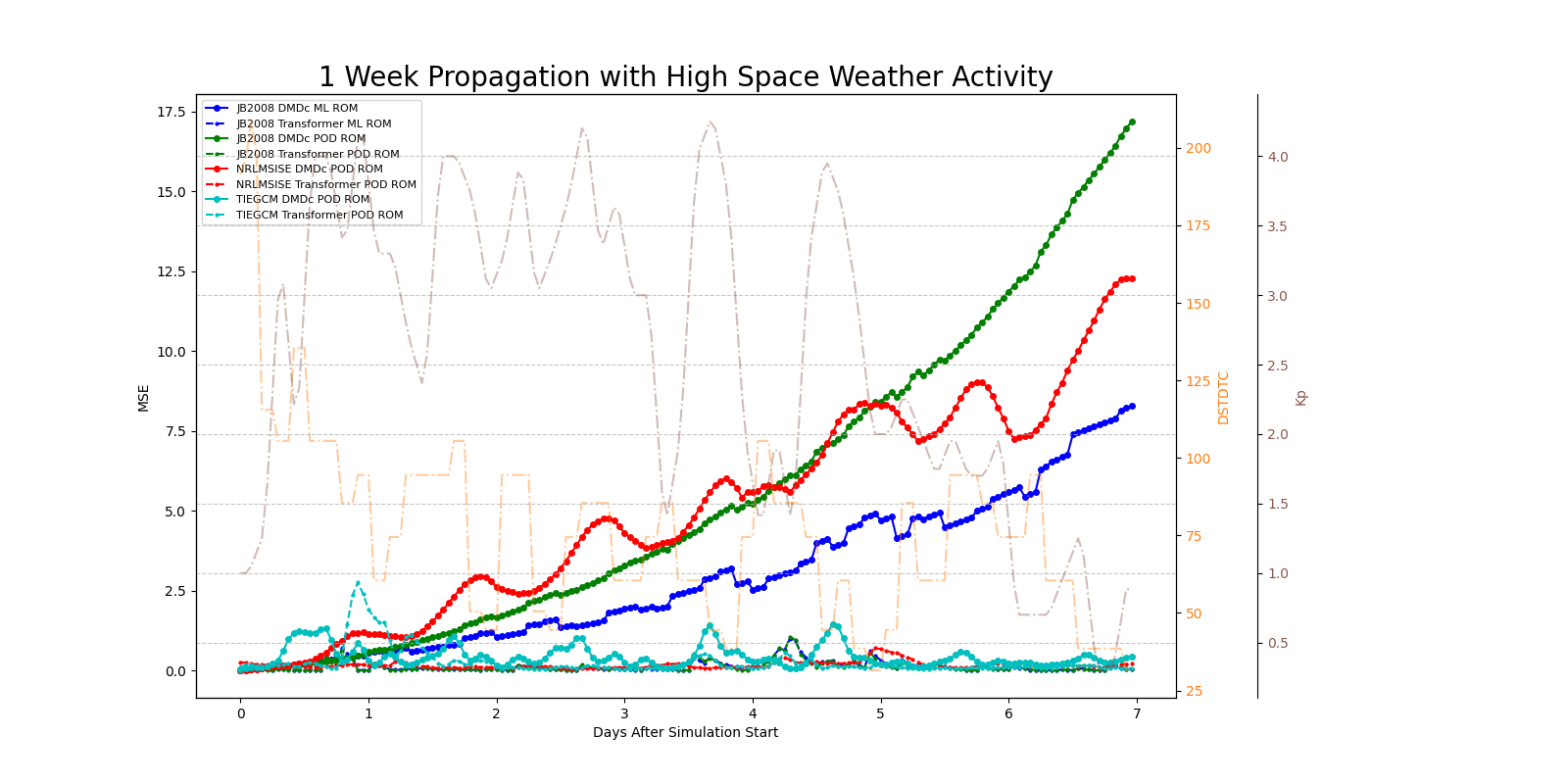}
\centering
\caption{MSE for one week propagation on high space weather activity test data.}
\label{fig:propagationhigh}
\end{figure}

The following plots show a comparison of DMDc and transformer-based propagation across all models for low solar activity (Figure \ref{fig:propagationlow}), medium solar activity (Figure \ref{fig:propagationmedium}), and high solar activity (Figure \ref{fig:propagationhigh}). Each hour in the forecast is represented by one data point.

In the low solar activity plot (Figure \ref{fig:propagationlow}), the JB2008 Transformer ML ROM maintains the lowest MSE, remaining less than 2 for the entire propagation period, closely followed by the JB2008 Transformer POD ROM. While the JB2008 transformer models remain relatively stable over the propagation period, the NRLMSISE Transformer POD ROM and TIEGCM Transformer POD ROM have larger spikes in MSE. On day 6, the NRLMSISE Transformer POD ROM reaches almost 5 MSE and the TIEGCM Transformer POD ROM spikes at over 11 MSE. Additionally, a small spike in error occurs for the JB2008 transformer models on the same day. Upon observing DSTDTC, there is a spike in geomagnetic activity occurring around the same time. While the transformer models all dominated their DMDc counterparts for each model, both the JB2008 DMDc ML ROM and JB2008 DMDc POD ROM maintain relatively low MSE, of less than 2, for the duration of the forecast. The TIEGCM DMDc POD ROM is shown to have a long spike in MSE from 1 to 3 days after simulation start, jumping from less than 2 MSE to up to 9 MSE. Finally, the NRLMSISE DMDc POD ROM demonstrates a consistent cyclic error increase over the propagation window, reaching over 13 MSE on the last day.

The medium solar activity plot (Figure \ref{fig:propagationmedium}) shows all models maintaining less than 2.5 MSE over the entire propagation window. Since this maximum error is much less than the 14 MSE maximum from the low solar activity plot, the errors in this plot appear to spike more. The NRLMSISE Transformer POD ROM maintains a consistent less than 0.25 MSE over the forecast window, with the JB2008 Transformer ML ROM maintaining a similar error range, except for the spike to almost 0.75 MSE before day 1. The JB2008 Transformer POD ROM and TIEGCM Transformer POD ROM show much larger spikes, ranging from 0.75-1.4 MSE. For the DMDc models, the NRLMSISE DMDc POD ROM, similarly to the NRLMSISE transformer, dominates the DMDc models in MSE for much of the forecast window. Both the JB2008 DMDc ML ROM and the JB2008 DMDc POD ROM grow in MSE almost monotonically, while the TIEGCM DMDc POD ROM fluctuates from 0 to 2.5 MSE over the entire forecast window.

Lastly, the high solar activity plot (Figure \ref{fig:propagationhigh}) illustrates differing performance between the JB2008 and NRLMSISE DMDc models and the TIEGCM DMDc model and transformer models. While the JB2008 DMDc ML ROM, JB2008 DMDc POD ROM, and the NRLMSISE DMDc POD ROM increase in MSE for the propagation period, the TIEGCM DMDc POD ROM, JB2008 Transformer ML ROM, JB2008 Transformer POD ROM, NRLMSISE Transformer POD ROM, and TIEGCM Transformer POD ROM maintain about constant MSE, of around 1.25 MSE over the propagation period. Compared to the low solar activity plot, it is apparent that the JB2008 DMDc propagated models do not maintain less than 2 MSE performance when applied in high solar activity conditions. TIEGCM DMDc POD ROM, the only physics-based reduced-order model in this study, is the only DMDc ROM which maintains less than 2 MSE at high space weather activity conditions. Interestingly, the TIEGCM DMDc POD ROM had large MSE spikes in low space weather conditions. Out of the dominating transformer models, the JB2008 Transformer POD ROM, JB2008 Transformer ML ROM, and NRLMSISE Transformer POD ROM exhibit comparable performance, of less than 1.25 MSE for the propagation window. The TIEGCM Transformer POD ROM had an error spike of over 2.5 MSE 1 day after simulation start, which occurred after a steep rise in the Kp index.

Overall, errors often accumulate quickly over the one week time period for the DMDc propagated ROMs, while only spiking at specific timesteps for the transformer-based ROMs. Additionally, the transformer forecasters consistently dominated the DMDc forecasting method for all levels of space weather activity. With the high solar activity test case showing the most significant difference between DMDc and the transformer-based approach; while the JB2008 DMDc models were comparable to all transformer-based approaches for low space weather activity, they were no longer within 2 MSE of the transformer models in high space weather conditions. In addition, the TIEGCM DMDc POD ROM was shown to be the only DMDc propagated model to perform with sub-2.5 MSE for high space weather conditions, but similar performance was not demonstrated by the model for low space weather conditions. Finally, all models maintained sub-2.5 MSE in medium space weather conditions. From observing these test cases, TIEGCM model errors are shown to change with increases in the Kp index and similar variations in the other density models occurred in relation to changes in the DSTDTC index.

The results for all DMDc and Transformer-based 1 week forecasts in the test datasets, for low, medium, and high solar activity, are shown in Tables \ref{tab:low_solar}-\ref{tab:high_solar}. All performance metrics are computed for the reduced-order model, instead of projecting back into the full-order to compute each metric. This allows the true training errors to be computed, independent of the low-dimensional projection errors, which are controlled by the reduced-order modeling approach (POD or ML).

In Table \ref{tab:low_solar}, the transformer-based models maintain the lowest MSE for 1 week forecasts across the entire low space weather test dataset. With the lowest error of 0.12215 MSE achieved by the JB2008 Transformer ML ROM. As in Figure \ref{fig:propagationlow}, the JB2008 DMDc ML ROM and JB2008 DMDc POD ROM have comparable errors (0.38346 MSE and 0.82024 MSE) to the transformer-based models for low space weather activity.

\begin{table}[h]
\centering
\begin{tabular}{|c|c|c|c|c|}
\hline
Model & Transformer MSE & DMDc MSE & Transformer MAE & DMDc MAE \\
\hline
JB2008 POD ROM & \textbf{0.22178} & 0.82024 & \textbf{0.33443} & 0.62294 \\
NRLMSISE POD ROM & \textbf{0.84399} & 7.40740 & \textbf{0.66705} & 1.88537 \\
TIEGCM POD ROM & \textbf{0.64757} & 1.77646 & \textbf{0.54225} & 0.99615 \\
JB2008 ML ROM & \textbf{0.12215} &  0.38346 & \textbf{0.25138} & 0.44345 \\
\hline
\end{tabular}
\caption{MSE and MAE for the transformer and DMDc propagator for low solar activity.}
\label{tab:low_solar}
\end{table}

Table \ref{tab:medium_solar} also shows the lowest MSE for transformer-based models during medium space weather activity. With the JB2008 Tranformer ML ROM maintaining the lowest MSE of 0.33400, closely followed by the NRLMSISE Transformer POD ROM. The JB2008 DMDc POD ROM and JB2008 DMDc ML ROM now show the largest DMDc-based model errors (8.95260 MSE and 3.58280 MSE).

\begin{table}[h]
\centering
\begin{tabular}{|c|c|c|c|c|}
\hline
Model & Transformer MSE & DMDc MSE & Transformer MAE & DMDc MAE \\
\hline
JB2008 POD ROM & \textbf{0.55943} &  8.95260 & \textbf{0.45291} & 1.98441 \\
NRLMSISE POD ROM & \textbf{0.44961} & 1.40329 & \textbf{0.48112} & 0.58854 \\
TIEGCM POD ROM & \textbf{0.65792} & 1.06309 & \textbf{0.48025} & 0.73588 \\
JB2008 ML ROM & \textbf{0.33400} & 3.58280 & \textbf{0.40303} & 1.43284 \\
\hline
\end{tabular}
\caption{MSE and MAE for the transformer and DMDc propagator for medium solar activity.}
\label{tab:medium_solar}
\end{table}

For high space weather activity, Table \ref{tab:high_solar} shows the transformer-based forecasters dominating all empirical DMDc-based models by a larger margin, by between 2.2 and 12.16 MSE. While all transformer models are very close in propagator error, the JB2008 Transformer POD ROM performs slightly better than all others, with about a 0.056 difference in MSE. While not dominating any of the transformer models, the physics-based TIEGCM DMDc POD ROM manages to achieve the closest forecasting error to the transformer models, with about a 0.234 MSE difference.

\begin{table}[h]
\centering
\begin{tabular}{|c|c|c|c|c|}
\hline
Model & Transformer MSE & DMDc MSE & Transformer MAE & DMDc MAE \\
\hline
JB2008 POD ROM & \textbf{0.36612} & 12.52918 & \textbf{0.34554} & 2.31920 \\
NRLMSISE POD ROM & \textbf{0.42239} & 2.63724 & \textbf{0.48266} & 1.10854 \\
TIEGCM POD ROM & \textbf{0.43682} & 0.67073 & \textbf{0.44843} & 0.60545 \\
JB2008 ML ROM & \textbf{0.42245} & 6.07747 & \textbf{0.40448} & 1.74487 \\
\hline
\end{tabular}
\caption{MSE and MAE for the transformer and DMDc propagator for high solar activity.}
\label{tab:high_solar}
\end{table}

The transformer-based forecaster outperforms DMDc for all levels of space weather activity for the JB2008 POD, NRLMSISE POD, and TIEGCM POD, and JB2008 ML ROM models. Additionally, the transformer MSE and MAE were consistent between models and levels of space weather activity, varying between 0 - 0.85 and reaching a maximum value of 0.84399 MSE for NRLMSISE POD ROM propagation during low solar activity. DMDc forecasting, on the other hand, proved to be very sensitive to the input space weather indices and the initial density snapshot used for propagation; MSE varied between 0.38346 and 12.52918, reaching a minimum for the JB2008 ML ROM propagation during low solar activity and a maximum for the JB2008 POD ROM propagation during high solar activity.
 
\section{Discussion}
\label{sec:Discussion}

From conducting propagation analyses and error computations for both the transformer forecaster and the DMDc forecaster for the JB2008 POD, NRLMSISE POD, TIEGCM POD, and JB2008 ML ROMs, it is apparent that transformer-based forecasting provides predictions which are more robust and almost exclusively more accurate than DMDc predictions. Likely, the transformer's robustness can be drawn from the PatchTST architecture of enabling a look-back window. Instead of using only the previous hour of atmospheric density data, as DMDc does to make a prediction, the transformer forecaster is able to use over a month of previous data (with a 42 day look-back window) to provide a density forecast. Additionally, forecasts for test data made by DMDc had very high variances. Since the matrices, $A$ and $B$, are trained by data outside of the test dataset, the test data may be outside of the linear regime in which DMDc can be applied. Leading to compounding forecasting errors in very short time periods.

For future work, the transformer forecaster will be augmented to include a graph neural network to learn the cross-channel relationships between space weather indices and atmospheric density. This formulation will allow the space weather control input to affect atmospheric density as a nonlinear dynamical system, while maintaining noise robustness from the channel-independent transformer architecture.

\section{Conclusion}
\label{sec:Conclusion}

A transformer forecaster for the JB2008, NRLMSISE, and TIEGCM atmospheric density models has been formulated in this work and compared with DMDc forecasting methods. The transformer-based forecasting approach has been shown to be both accurate and robust when handling long-term dependencies in atmospheric density data, when compared to the linear DMDc forecaster. Importantly, this effectiveness extends across varying levels of solar activity. With the significant potential for a single geomagnetic storm to alter the orbits of Resident Space Objects (RSOs), our developed nonlinear transformer-based architecture for atmospheric density forecasting presents a significant advancement over traditional linear propagation methods.

\section*{Acknowledgement}
Research was sponsored by the United States Air Force Research Laboratory and the Department of the Air Force Artificial Intelligence Accelerator and was accomplished under Cooperative Agreement Number FA8750-19-2-1000. The views and conclusions contained in this document are those of the authors and should not be interpreted as representing the official policies, either expressed or implied, of the Department of the Air Force or the U.S. Government. The U.S. Government is authorized to reproduce and distribute reprints for Government purposes notwithstanding any copyright notation herein. This work was supported by a NASA Space Technology Graduate Research Opportunity and the National Science Foundation under award NSF-PHY-2028125. The authors would like to thank the MIT SuperCloud and Lincoln Laboratory Supercomputing Center for providing HPC, database, and consultation resources that have contributed to the research results reported in this paper. The authors would like to express their gratitude to Nicolette Clark for her substantial contributions to atmospheric density model development and analysis. Specifically, Clark's work on the development and uncertainty quantification of Reduced-Order Atmospheric Density Models (ROMs) for fast and accurate orbit propagation were essential for developing this work \cite{clark2023reduced}.

\section*{References}

\bibliographystyle{unsrt}

\begingroup
\renewcommand{\section}[2]{}%
\bibliography{references}

\begin{thebibliography}{10}

\bibitem{wu2021}
Ziyou Wu, Steven~L. Brunton, and Shai Revzen.
\newblock Challenges in dynamic mode decomposition.
\newblock {\em Journal of The Royal Society Interface}, 2021.

\bibitem{Turner2020}
Turner H., Zhang M., D.~Gondelach, , and R.~Linares.
\newblock {\em Machine Learning Algorithms for Improved Thermospheric Density Modeling}, volume 12312.
\newblock Springer, Cham, 2020.

\bibitem{wen2023}
Qingsong Wen, Tian Zhou, Chaoli Zhang, Weiqi Chen, Ziqing Ma, Junchi Yan, and Liang Sun.
\newblock Transformers in time series: A survey.
\newblock {\em arXiv preprint arXiv:2202.07125}, 2023.
\newblock Accepted by 32nd International Joint Conference on Artificial Intelligence (IJCAI 2023).

\bibitem{briden2022}
Julia Briden, Nicolette Clark, Peng~Mun Siew, Richard Linares, and Tzu-Wei Fang.
\newblock Impact of space weather on space assets and satellite launches.
\newblock In {\em 23rd Advanced Maui Optical and Space Surveillance Technologies}. Advanced Maui Optical and Space Surveillance Technologies, 2022.

\bibitem{clark2023reduced}
Nicolette~LeAnn Clark.
\newblock {\em Reduced-Order Atmospheric Density Modeling for LEO Satellite Orbital Reentry Prediction}.
\newblock PhD thesis, Massachusetts Institute of Technology, 6 2023.

\bibitem{Mehta2017}
Piyush~M. Mehta and Richard Linares.
\newblock A methodology for reduced order modeling and calibration of the upper atmosphere.
\newblock {\em Space Weather}, 15(10):1270--1287, 2017.

\bibitem{Loffe2015}
Sergey Ioffe and Christian Szegedy.
\newblock Batch normalization: Accelerating deep network training by reducing internal covariate shift.
\newblock {\em Proceedings of the 32nd International Conference on Machine Learning}, PMLR 37:448-456, 2015.

\bibitem{DeCarlo1989}
R.~A. DeCarlo.
\newblock {\em Linear systems: A state variable approach with numerical implementation}.
\newblock Upper Saddle River, NJ, USA: Prentice-Hall, Inc, 1989.

\bibitem{Goodfellow2016}
Ian Goodfellow, Yoshua Bengio, and Aaron Courville.
\newblock {\em Deep Learning}.
\newblock MIT Press, 2016.
\newblock \url{http://www.deeplearningbook.org}.

\bibitem{nie2023}
Yuqi Nie, Nam~H. Nguyen, Phanwadee Sinthong, and Jayant Kalagnanam.
\newblock A time series is worth 64 words: Long-term forecasting with transformers.
\newblock In {\em International Conference on Learning Representations (ICLR)}, 2023.

\bibitem{tsai}
Ignacio Oguiza.
\newblock tsai - a state-of-the-art deep learning library for time series and sequential data.
\newblock Github, 2022.

\bibitem{Gondelach2020}
David~J. Gondelach and Richard Linares.
\newblock {Real-Time Thermospheric Density Estimation via Two-Line Element Data Assimilation}.
\newblock {\em Space Weather}, 18(2):e2019SW002356, 2020.

\end{thebibliography}
\endgroup

\end{document}